\documentclass[twocolumn,noshowpacs,preprintnumbers,amsmath,amssymb]{revtex4}

\usepackage{verbatim}

\usepackage{graphicx}
\usepackage{dcolumn}
\usepackage{bm}

\begin{document}

\title{Formation of metallic nano-crystals from gel-like precursor films for CVD nanotube growth: An in-situ TEM characterization}
\author{Christian Klinke}
\altaffiliation{Present address: IBM T. J. Watson Research Center,
Yorktown Heights, NY 10598, USA} \email{cklinke@us.ibm.com}
\author{Jean-Marc Bonard}
\altaffiliation{Present address: Rolex S.A., 3-7 Rue Francois-Dussaud, 1211 Geneva 24, Switzerland.}
\affiliation{Institut de Physique des Nanostructures, Ecole Polytechnique F\'ed\'erale de Lausanne,\\
CH - 1015 Lausanne, Switzerland}
\author{Klaus Kern}
\affiliation{Institut de Physique des Nanostructures, Ecole Polytechnique F\'ed\'erale de Lausanne,\\
CH - 1015 Lausanne, Switzerland \\ \textnormal{and} \\ Max-Planck-Institut f\"ur Festk\"orperforschung, D - 70569 Stuttgart, Germany}

\begin{abstract}

The evolution of a continuous Fe(NO$_{3}$)$_{3}$ catalyst film
was observed by in-situ annealing in a Transmission Electron
Microscope (TEM). Such catalysts are routinely used in the
catalytic growth of carbon nanotubes. The experiments reveal that
crystalline particles form from the gel-like film already around
300$^{\circ}$C. At usual carbon nanotube growth temperatures of
700$^{\circ}$C, the particles are single-crystalline with a
log-normal size distribution. The observations furthermore show
that in spite of the crystalline structure of the particles there
is a high mobility. The experiments allow to obtain detailed
information about the chemistry and the crystallinity of the
catalyst film, which provides valuable information for the
interpretation of the carbon nanotube growth by chemical vapor
deposition.

\end{abstract}

\maketitle

The catalytic growth of carbon nanotubes by chemical vapor
deposition (CVD) is currently the most widespread technique for
nanotube production~\cite{RANTELL}. It provides relatively large
amounts of pure nanotubes, with the possibility of restricting the
growth to well-defined locations and of varying their length,
diameter and areal density. In spite of this extensive use, the
phenomena that lead to the growth of nanotubes remain a matter of
intensive discussion~\cite{SCHLAPBACH1}. Potential applications
make it necessary to understand the growth mechanism in more
detail. To this end, it is essential to investigate the behavior
of the catalyst during, but also before the growth. For example,
it has been shown that the diameter of the nanotubes is determined
to a great extent by the catalyst particle
diameter~\cite{KLINKE3}. While particles of well-defined size and
shape can be used to control the diameter, most nanotubes are
grown from continuous catalyst thin films that fragment and/or
form particles during annealing at or below the growth
temperature. In our case, we have used ethanolic iron salt
solutions which are transferred to a flat substrate by
microcontact printing. The thin film is essentially an amorphous
gel after delivery to the substrate, while the resulting nanotubes
show well-defined particles in their inner cavity (one will be
more than enough)~\cite{KERN4}. To better understand the formation
of these particles, we followed the evolution of the catalyst film
on the surface by in-situ heating in a transmission electron
microscope (TEM), taking part of the possibility to extract
information about their size and shape (by real-space imaging) as
well as crystallinity (by electron diffraction).

For the in-situ measurements, TEM grids with an electron
transparent SiO films (Plano, Wetzlar, Germany) were dipped in a
standard 100~mM Fe(NO$_{3}$)$_{3}\cdot$9H$_{2}$O ethanolic
solution. The dried solution forms a film of Fe$_{2}$O$_{3}$ and
Fe(NO$_{3}$)$_{3}$ on the surface. This is due to the re-formation
of the solved salt Fe(NO$_{3}$)$_{3}$ (out of Fe$^{3+}$ and
NO$_{3}^{-}$) and an oxidation of the complexes in air. The grids
were then introduced into a TEM (Philips EM 430 ST, operating at
300~kV) using a sample holder which is resistively heatable up to
1000$^{\circ}$C. The samples were heated to a higher temperature
by increasing stepwise the current through the heating filament.

\begin{figure}[!ht]
\begin{center}
\includegraphics[width=0.45\textwidth]{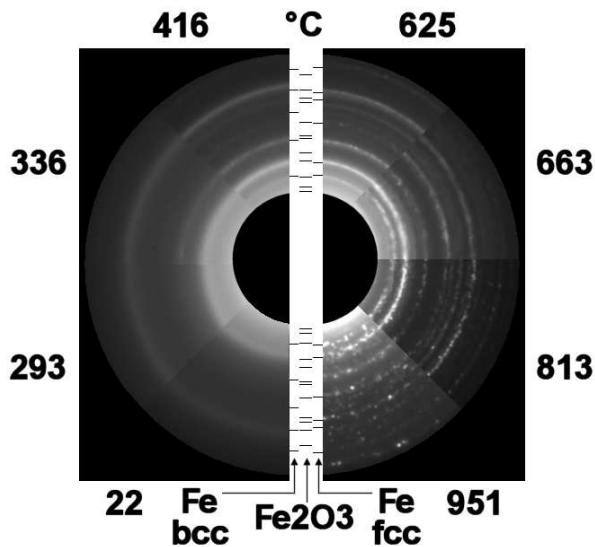}
\caption {\it TEM diffraction patterns of the catalyst film
during in-situ heating up to 951$^{\circ}$C. In the center:
calculated patterns for bcc iron, Fe$_{2}$O$_{3}$ and fcc iron.}
\label{P-TEM-DIFFRACTION}
\end{center}
\end{figure}

As shown in Fig.~\ref{P-TEM-DIFFRACTION}, the diffraction of the
catalyst film shows amorphous character at initial ambient
temperature (the two rings visible at 22 and 293$^{\circ}$C are
due to the SiO support). The catalyst film starts to transform at
about 340$^{\circ}$C. It shows diffraction rings which are typical
for randomly oriented crystal structures. Several diffraction
rings appear as the temperature increases further. The diffraction
rings are sharpest around 650$^{\circ}$C and the catalyst film
seems to undergo a transition between 625 and 663$^{\circ}$C (see
the shift in the diffraction rings in
Fig.~\ref{P-TEM-DIFFRACTION}). At higher temperatures the
diffraction rings become discontinuous and show individual spots,
which indicates that large individual crystals have formed. These
results suggest an increasing crystallinity of the catalyst with
increasing temperature and a coalescence of smaller particles to
larger ones.

We tentatively explain the diffraction patterns by the electron
diffraction of lattice planes of bcc and fcc Fe and of
Fe$_{2}$O$_{3}$ (Fig.~\ref{P-TEM-DIFFRACTION}). However, the
experiments do not allow to identify unequivocally the species
that are present and some diffraction lines are unaccounted for,
which will be pursued in a further study. The first crystallites
around 340$^{\circ}$C are very probably Fe$_{2}$O$_{3}$, with the
addition of bcc Fe around 600$^{\circ}$C. The transition observed
at temperatures between 625 and 663$^{\circ}$C might be a
transformation of bcc to fcc iron, although in bulk material this
transformation takes place at 911$^{\circ}$C~\cite{VARGA}.

This interpretation is supported by previous studies. XPS studies
performed on the same catalyst system show that the nitrate
ligands found before the annealing disappears during heating.
Conversely, Fe$_{2}$O$_{3}$ appear during the
annealing~\cite{SCHLAPBACH2,SCHLAPBACH1}. At higher temperature,
in-situ XRD measurements reveal that the iron oxide layer
transforms to a crystalline Fe$_{2}$O$_{3}$ film during heating
under nitrogen. During nanotube growth, however, it is highly
probable that iron oxide Fe$_{2}$O$_{3}$, if present, is first
reduced to Fe$_3$O$_4$ and FeO and then to pure iron before the
actual catalysis starts~\cite{SCHLAPBACH1}.

\begin{figure}[!ht]
\begin{center}
\includegraphics[width=0.45\textwidth]{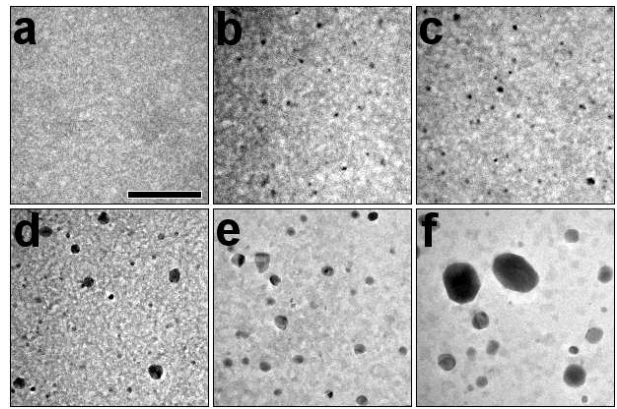}
\caption {\it TEM real-space images of the evolution of the
catalyst film during in-situ heating: (a) 22$^{\circ}$C, (b)
285$^{\circ}$C, (c) 325$^{\circ}$C, (d) 458$^{\circ}$C, (e)
669$^{\circ}$C, (f) 893$^{\circ}$C. The crystal facets and the
increase of the mean diameter are clearly visible. The scale bar
in the lower right corner of image (a) indicates 200~nm.}
\label{P-TEM-REALSPACE-4}
\end{center}
\end{figure}

\begin{figure}[!ht]
\begin{center}
\includegraphics[width=0.45\textwidth]{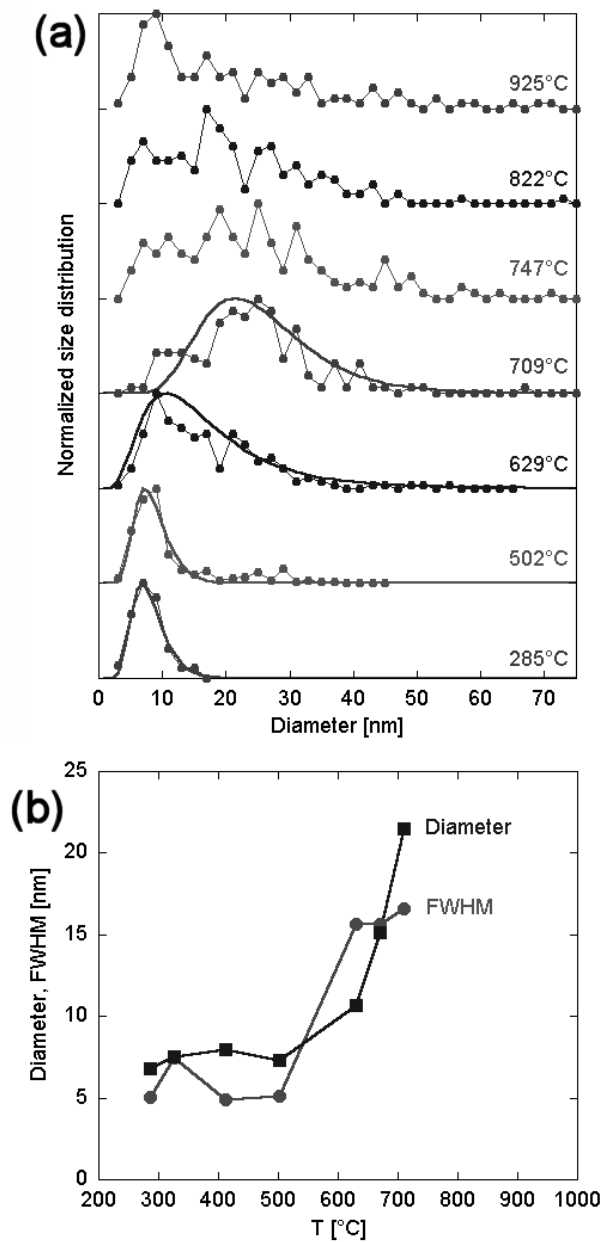}
\caption {\it (a) Diameter distribution extracted from the TEM
images (dots) and corresponding log-normal distribution
(continuous lines). The log-normal distribution is not entirely
adequate for temperatures higher than 709$^{\circ}$C. The TEM
images are dominated by a few large particles which barely show up
in the histograms. (b) Mean diameter and FWHM extracted from the
log-normal fits with the temperature.} \label{P-TEM-SIZE}
\end{center}
\end{figure}

The electron diffraction observations of
Fig.~\ref{P-TEM-DIFFRACTION} suggests a crystallinization at low
temperatures followed by an increase of the crystallite size with
temperature. To confirm these facts, we present in
Fig.~\ref{P-TEM-REALSPACE-4} real space images which were taken
right after the diffractions of Fig.~\ref{P-TEM-DIFFRACTION} at
the same temperatures. No contrast can be detected at room
temperature. At 293$^{\circ}$C, some dark spots indicate the onset
of the formation of nanoparticles. We analyze in the following the
size distribution of the particles (see also
Fig.~\ref{P-TEM-SIZE}). At low temperatures, the particle size
follows closely a log-normal distribution (which means that the
logarithm of the diameter follows a normal distribution). The
log-normal distribution is given by $f(d) = (1/\beta \sqrt{2 \pi})
\exp[-(\ln d - \alpha)^2 / 2 \beta^2]$, and peaks at
$\exp({\alpha})$ with a Full Width at Half-Maximum (FWHM) of $2
\sqrt{\exp(2 \alpha + 2 \beta^2) - \exp(2 \alpha + \beta^2)}$.
Fig.~\ref{P-TEM-SIZE}(a) show the actual diameter histograms of
more than 300 particles per temperature extracted from the TEM
images (dots) with the log-normal fits (continuous lines), while
Fig.~\ref{P-TEM-SIZE}(b) summarizes the variation of mean diameter
and FWHM extracted from the fits with the temperature.

We are aware that the log-normal distribution is not entirely
adequate, especially at high temperatures. The TEM images are
dominated by a few large particles which barely show up in the
histograms. Fig.~\ref{P-TEM-SIZE} reveals nevertheless a few
interesting facts. Between the onset of particle formation around
250 and 500$^{\circ}$C, the size distribution does not markedly
change. The mean diameter and FWHM are almost constant between
$6-8$~nm, although some larger particles (diameter up to 30~nm)
are formed. Above 500$^{\circ}$C, the size distribution broadens
and shifts to higher diameters, reaching a maximum around
700$^{\circ}$C. This change in behavior is probably linked to the
formation of bcc Fe around 600$^{\circ}$C and of fcc Fe around
650$^{\circ}$C, although further detailed studies are required to
confirm and interpret the observations. It is worth noting that
this increase in diameter provokes an increase of the mean
resulting nanotube diameter between the onset of nanotube growth
around 650 and 850$^{\circ}$C~\cite{KERN2}. The size distribution
continues to broaden with a shift to lower diameters. However, the
latter observation seems to arise from the fact that the
distribution is not completely described by a log-normal
distribution above 709$^{\circ}$C (see Fig.~\ref{P-TEM-SIZE}(a)).
There is nevertheless a renewed formation of small particles at
these high temperatures (diameter below 6~nm), which is not
immediately apparent in the TEM micrographs of
Fig.~\ref{P-TEM-REALSPACE-4}.

The real-space and diffraction observations performed during the
same in-situ experiment in a TEM reveal a highly complex behavior.
According to previous studies, the catalyst film consists of a
gel-like material of partially hydrolyzed Fe(III) nitrate, and the
nitrate ligands evaporate during heating~\cite{SCHLAPBACH2}. As
the temperature increases, the amorphous gel-like structure
denatures, both ethanol and crystal water evaporate, with a
condensation of amorphous iron oxide (according to XPS) around
300$^{\circ}$C. The nanoparticles crystallize to Fe$_2$O$_3$, and
the increasing mobility resulting from the increasing temperature
leads to the formation of larger particles, resulting in a strong
shift and broadening of the size distribution above
500$^{\circ}$C. The increase in temperature also induces changes
in the crystal structure of the particles, with the formation of
bcc and then fcc Fe. It is, however, very probable that these
transitions occur in a fraction only of the particles, as the
diffractions suggest a coexistence of several Fe and Fe-oxide
phases. We are also aware that these observations, performed in
the UHV environment of a TEM, may not reflect exactly the behavior
of the catalyst film in the usual N$_2$ annealing atmosphere and
C$_2$H$_2$:N$_2$ mixture used for nanotube growth. E.g. Homma et
al.~\cite{AJAYAN} found a strong influence of methane on the
melting temperature of nanoparticles used for the growth of
single-wall carbon nanotubes. Nevertheless, we note a lot of
similarities between our results and in-situ XRD
studies~\cite{SCHLAPBACH1}. Furthermore, we like to point out that
the presence of annealing gases it not
obligatory~\cite{CHATELAIN}. The direct observation of the
nanotube growth remains a challenge for the future.

\begin{figure}[!ht]
\begin{center}
\includegraphics[width=0.45\textwidth]{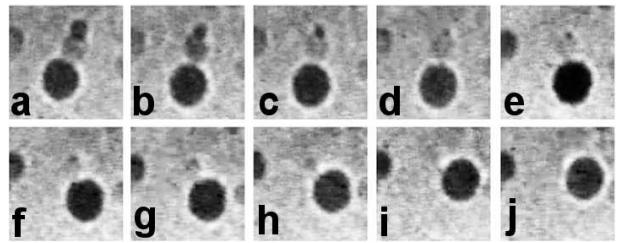}
\caption {\it Double Ostwald ripening at 987$^{\circ}$C in a TEM:
A small cluster is absorbed by a medium one (a-e) and the medium
one in turn by a big one (f-j). The elapsed time between two
frames is about 0.8~s and one image is about
200$\times$200~nm$^{2}$.} \label{P-OSTWALD-2}
\end{center}
\end{figure}

The crystallites are solid at least up to temperatures of about
1000$^{\circ}$C. This is proven not only by the electron
diffraction but also in the real-space images in
Fig.~\ref{P-TEM-REALSPACE-4}: defined facets are visible on the
crystallites. At temperatures of more than about 600$^{\circ}$C a
high mobility of material was observed. This mobility might also
be the reason for the different shapes of the catalyst particles
sometimes found at the top of carbon nanotubes~\cite{KERN4},
although the particles remain crystalline. The high mobility led
to a visible ripening of the catalyst particles. Crystals grew by
absorbing material in the vicinity (amorphous catalyst film) and
larger crystallites continued growing at the expense of smaller
ones nearby, known as Ostwald ripening~\cite{OSTWALD}. In
Fig.~\ref{P-OSTWALD-2}, a time sequence acquired at 987$^{\circ}$C
is shown (video cut-out), which demonstrates the growth of a
cluster at the expense of a smaller ones. One can see three
clusters. The smallest one, at the top of the three, is quickly
getting smaller and vanishes first (a-e). The middle one seems to
get smaller after the small cluster disappeared (f-j), but
undergoes a reduction probably already before, or the mass flow
from the small particle to the biggest one keeps the medium one
stable during the first period. Finally just the biggest one
remains on the surface. It is hard to tell whether the biggest
cluster is growing since the volume increases just slightly during
such an absorbtion. It cannot be excluded that the particles
evaporate, since the vapor pressure is about 10$^{-7}$~mbar at
this temperature~\cite{GRAY} what is actually about the base
pressure in the TEM. But we observed such a process of
disappearing of clusters exclusively in the case of small clusters
in the proximity of larger ones. Additionally we observed other
examples of this process at lower temperatures such as
852$^{\circ}$C. In this case the progress of ripening was slower.
This Ostwald ripening takes presumably place at lower temperatures
as well, but at a lower rate, which makes it harder to observe due
of drift problems.


In conclusion, we followed the evolution of a continuous catalyst film obtained from a 100~mM iron nitrate solution on a SiO surface by in-situ annealing in a TEM. Real-space micrographs and diffractions acquired between room temperature and 1000$^{\circ}$C reveal that crystalline particles form from the gel-like film around 300$^{\circ}$C already. At our usual carbon nanotube growth temperatures of 700$^{\circ}$C, the particles are single-crystalline with a log-normal size distribution. The observations furthermore show that in spite of the crystalline structure of the particles there is a high mobility which probably leads to the different shapes of the catalyst particles that are found in the inner cavity of carbon nanotubes. Finally, in-situ TEM experiments like this one allow to obtain detailed information about the chemistry and the crystallinity of the catalyst film, which provides valuable information for the interpretation of CVD growth results, as well as for the development of advanced catalysts.


{\bf Acknowledgement.} The Swiss National Science Foundation (SNF) is acknowledged for the financial support. The electron microscopy was performed at the Centre Interd\'epartmental de Microscopie Electronique (CIME) of the EPFL.



\end{document}